\documentclass[preprint,aps,prc,showpacs,nofootinbib]{revtex4}
\usepackage{epsfig}
\usepackage{graphicx,amsmath}

\newcommand{\Li}{\mbox{Li}_2}

\newcommand{\dd}{\mbox{d}}

\newcommand\ba{\begin{eqnarray}}
\newcommand\ea{\end{eqnarray}}
\newcommand\be{\begin{equation}}
\newcommand\ee{\end{equation}}
\newcommand\nn{\nonumber}

\def\Li#1#2{{\mathrm{Li}}_{#1}\left(#2\right)}

\begin{document}

\title{Double logarithmical corrections to beam asymmetry in polarized
electron-proton scattering}

\affiliation{\it JINR-BLTP, 141980 Dubna, Moscow region, Russian Federation}
\author{E. Kuraev}
\affiliation{\it JINR-BLTP, 141980 Dubna, Moscow region, Russian Federation}
\author{S.~Bakmaev}
\affiliation{\it JINR-BLTP, 141980 Dubna, Moscow region, Russian Federation}
\author{V.~V.~Bytev}
\affiliation{\it JINR-BLTP, 141980 Dubna, Moscow region, Russian Federation}
\author{Yu.~M.~Bystritskiy}
\affiliation{\it JINR-BLTP, 141980 Dubna, Moscow region, Russian Federation}

\author{E.~Tomasi-Gustafsson}
\affiliation{\it DAPNIA/SPhN, CEA/Saclay, 91191 Gif-sur-Yvette
Cedex, France }

\date{\today}

\begin{abstract}
The up-down asymmetry in transversally polarized electron proton scattering is induced by the interference between one and two photon exchange amplitudes. Inelastic intermediate hadronic states (different from one-proton state) of the two photon exchange amplitude give rise to contributions containing the square of "large logarithm" (logarithm of the ratio
of the transferred momentum to the electron mass).
We investigate the presence of such contributions in higher orders of
perturbation theory.
The relation with the case of zero transfer momentum is explicitly given.
The mechanism of cancellation of infrared singularities is discussed.
\end{abstract}
\maketitle
\section{Introduction}

\begin{figure}
\includegraphics[scale=1.0]{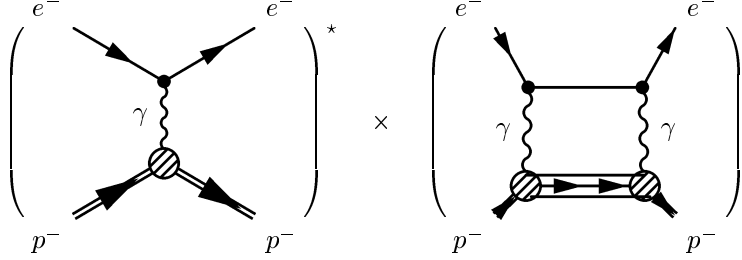}
\caption{The beam asymmetry due to transversal polarization of an electron beam in lowest order}
\label{fig_asym_LO}
\end{figure}

The beam asymmetry due to transversal polarization of an electron
beam scattered on (unpolarized) protons is a pure quantum effect
arising from the interference of the Born amplitude (with one
photon exchange) and the imaginary part of the two-photon exchange
amplitude (Fig. \ref{fig_asym_LO}).

The corresponding contribution to the differential cross section
as well as to the beam asymmetry is proportional to the electron mass.
Therefore, the presence of this contribution does not contradict the Kinoshita-Lee-Nauenberg theorem \cite{KLN}, about cancellation of mass singularities, since the corresponding cross sections are suppressed by the lepton mass.

We show that the main
contribution arises from the kinematical region of loop momenta when the energy of the electron in
the intermediate state, being on mass shell, is small in the reference frame of the center of mass
of the initial particles.

The kinematics is determined by the conservation laws \ba
e^-(a,p_1)+P(p)+(\gamma)\to e^-(p_1'')+X+(\gamma) \to
e(p_1')+P(p')+(\gamma) \ea where $a$ is electron spin. In general,
the emission of real photons must be considered, in order to avoid
infrared divergences, when  higher orders of perturbation theory
(PT) are taken into account. For the case of one proton
intermediate state in the two-photon exchange amplitude (TPE) the
energies  of the initial intermediate state and the final
elastically scattered electrons are equal
\begin{gather}
p_{10}=p''_{10}=p'_{10}=\epsilon=\frac{s-M^2}{2\sqrt{s}},\quad s=(p_1+p)^2, \nn \\
t=(p_1-p_1')^2=-2\epsilon^2(1-c),\quad p_1^2=(p_1')^2=m^2,\quad p^2=(p')^2=M^2,
\end{gather}
where $c=\cos\theta$ and $\theta$ is the scattering angle in the center of mass frame. The absolute values of the photon momenta squared, in TPE amplitude, can reach zero:
\begin{gather}
\Big |t_1\Big |=\Big |(p_1-p_1'')^2\Big |=2\epsilon^2(1-c_1),\quad \Big |t_2\Big |=\Big |(p'_1-p_1'')^2\Big |=2\epsilon^2(1-c_2),
\end{gather}
where $c_{1,2}$  are the cosines of the angles between the initial and intermediate electron momenta
and the intermediate and final ones, respectively.

For the case of inelastic hadronic intermediate state (for instance a
nucleon and a pion) the energy of the electron in the intermediate state $\epsilon''$ does not
exceed $\epsilon: m<\epsilon''<\epsilon $. The exchanged photon momenta squared become:
\ba
t_{1,2}=-2\epsilon\epsilon''(1-bc_{1,2}), \quad
1-b^2=\frac{m^2}{(\epsilon'')^2}(1-\frac{\epsilon''}{\epsilon})^2.
\ea
The main contribution arises from two regions $-t_1\ll t_2=t$ and $-t_2\ll t_1=t$.

Moreover, we will show that the energy of the intermediate electron is much
lower than the electron energy corresponding to the elastic case $\epsilon''\ll\epsilon$.

Neglecting the dependence on $p_1''$ from the remaining part of
the amplitude we find the main (double-logarithmical, or DL)
asymptotic behavior of the amplitude: \ba
(M_{box}^*M_{born})^{(DL)}\approx\int\frac{\epsilon''\dd\epsilon''
\dd O''}{2\pi t_1t_2}\approx \frac{-1}{4t}{L}^2. \ea We do not
distinguish here the two kinds of "large logarithms" \ba
L_s=\ln\frac{s}{m^2}-i\pi,\quad
L_t=\ln\frac{-t}{m^2}=\ln\frac{2\epsilon^2(1-c)}{m^2}. \ea The
result of exact calculation consists in the replacement \ba
{L}^2\to L_tL_s.\ea

\section{Calculation of Double Logarithm correction}

\begin{figure}
\includegraphics[scale=1.0]{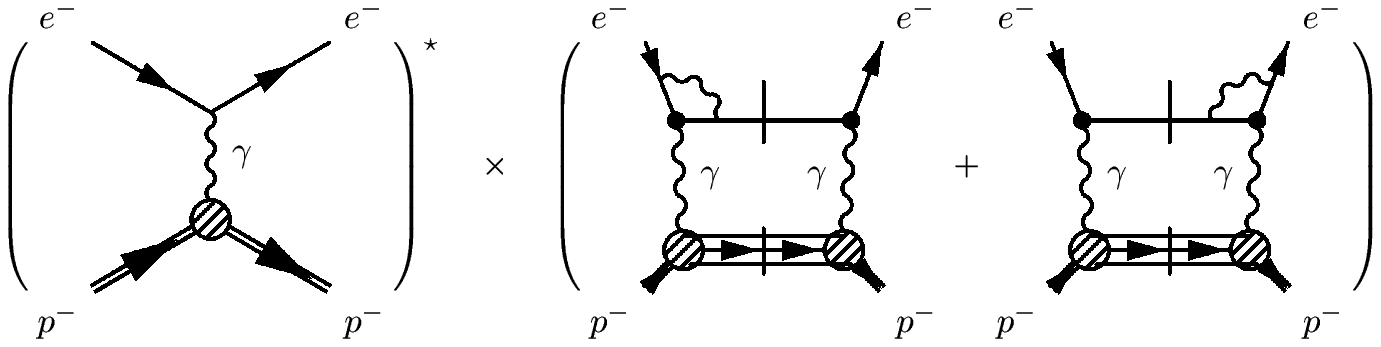}
\includegraphics[scale=1.0]{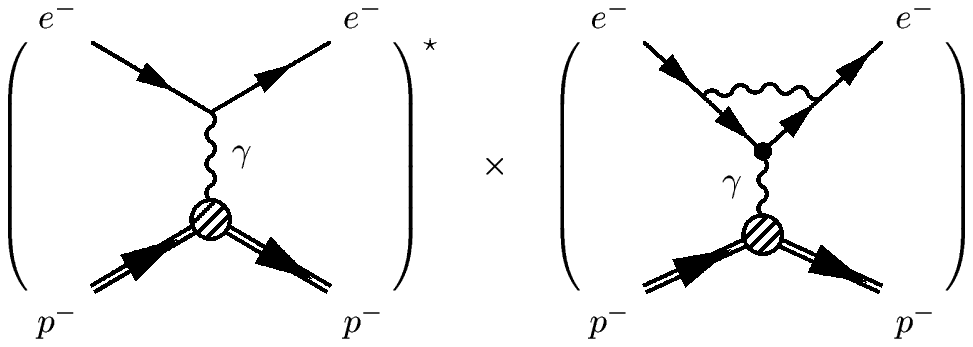}
\caption{Vertex corrections  $\delta_v$ which contribute to the asymmetry}
\label{corr_type_1}
\end{figure}

\begin{figure}
\includegraphics[scale=1.]{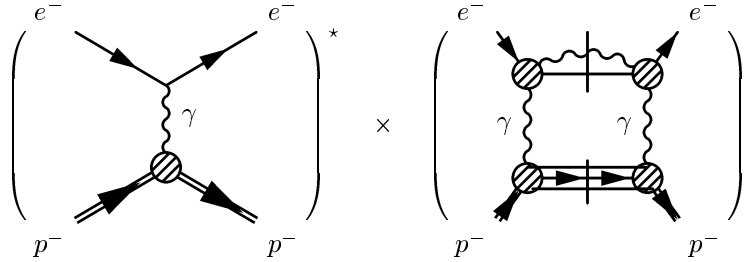}
\caption{Inner bremsstrahlung in two photon diagram}
\label{fig_virt_soft}
\end{figure}

\begin{figure}
\includegraphics[scale=1.]{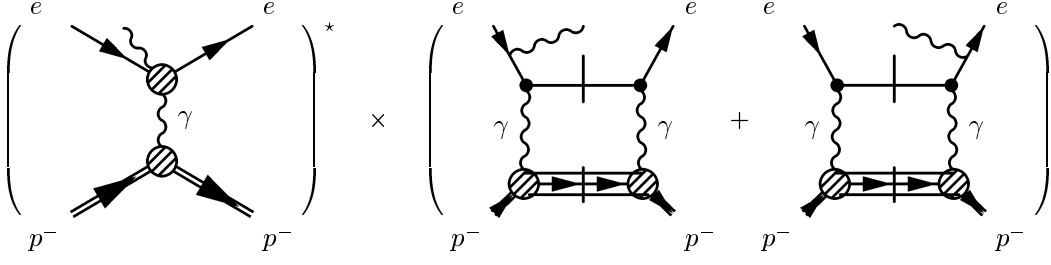}
\caption{Inelastic two photon contribution $\delta^{inel}_s$}
\label{fig_soft}
\end{figure}

Let us investigate the question about the size of radiative
corrections to the cross section for the scattering of
transversally polarized electron and to the relevant beam
asymmetry. The corrections in lowest order can be of several types
(see figs. \ref{corr_type_1}, \ref{fig_virt_soft}, \ref{fig_soft},
\ref{fig_sq_box}, \ref{tr_phot}).

Let us consider first the radiative corrections (RC) at the lowest
order. The contribution from the emission of virtual photons is
twofold. Firstly,  it is due to the vertex functions in the
kinematics where both electrons are on mass shell and the photon
mass squared is negative and large (in absolute value) compared to
the electron mass squared (see Fig. \ref{corr_type_1}). The main
contribution arises from the Dirac form factor. In the scattering
channel we have:
\begin{gather}
F_1(q^2)=1+\frac{\alpha}{\pi}\biggl[\ln\frac{m}{\lambda}(1-L_q)-1+\frac{3}{4}L_q-\frac{1}{4}L_q^2+
\frac{\pi^2}{12}\biggr]=1+\frac{\alpha}{\pi}\delta_v(q^2), \nn\\
 q^2=t, t_1, t_2, \quad L_q=L_t,L_1,L_2,\quad  L_1=\ln\frac{-t_1}{m^2},\quad L_2=\ln\frac{-t_2}{m^2}.
\end{gather}
The second class of contributions arise from emission and absorption of real photons as an intermediate state of the leptonic block.

Let firstly restrict our considerations to the emission of soft
real intermediate photons (see Fig. \ref{fig_virt_soft}).

For the corresponding contribution we have
\ba
\frac{\alpha}{\pi}\delta_s=-\frac{4\pi\alpha}{16\pi^3}\int\frac{\dd^3k}{\omega}\left (-\frac{p_1}{p_1k}+
\frac{p_1''}{p_1''k}\right )\left (\frac{p''_1}{p''_1k}-
\frac{p_1'}{p_1'k}\right )\Big |_{\omega<\Delta_1},\quad \Delta_1\ll\epsilon.
\ea
Using the expressions
\ba
-\frac{4\pi\alpha}{16\pi^3}\int\frac{\dd^3k}{\omega}\frac{m^2}{(p_1''k)^2}\Big |_{\omega<\Delta_1}
&=&
-\frac{\alpha}{\pi}\ln\frac{m\Delta_1}{\lambda\epsilon''}, \nn \\
\frac{4\pi\alpha}{16\pi^3}\int\frac{\dd^3k}{\omega}\frac{2p_1p_1''}{(p_1k)(p_1''k)}\Big |_{\omega<\Delta_1}
&=&
 \nn \\
\frac{\alpha}{2\pi}\left[ L_1\ln(\frac{m^2\Delta_1^2}{\lambda^2\epsilon\epsilon''})
\right .&+&\left .
\frac{1}{2}L_1^2-\frac{1}{2}\ln^2 \left (\frac{\epsilon''}{\epsilon}\right )-\frac{\pi^2}{3}+\Li{2}{\frac{1+c_1}{2}}\right ],
\ea
where $\lambda$ is a fictitious "photon mass", the resulting contribution
\ba
\delta_{vs}=\delta_v(t)+\delta_v(t_1)+\delta_v(t_2)+\delta_s
\ea
suffers from infrared divergences. To remove these divergencies, we must take into account
the inelastic process of electron-proton scattering with emission of additional
soft (or hard) real photons by initial and final electrons (see Fig \ref{fig_soft}):
\ba
\frac{\alpha}{\pi}\delta_s^{inel}=-\frac{4\pi\alpha}{16\pi^3}\int\frac{\dd^3k}{\omega}
\left (-\frac{p_1}{p_1k}+\frac{p_1'}{p_1'k}\right )^2\Big |_{\omega<\Delta_2},\quad \Delta_2\ll\epsilon.
\ea

The total sum $\delta=\delta_{vs}+\delta^{inel}_s$ is free from
infrared divergences: \ba
\delta&=&2(L_t-1)\ln\frac{\Delta_2}{\epsilon}+\frac{1}{2}(L_1+L_2)\ln\frac{\Delta_1^2}{\epsilon\epsilon''}
-\ln\frac{\Delta_1}{\epsilon''}-L_t\ln\frac{\Delta_1}{\epsilon}+
\frac{3}{4}(L_t+L_1+L_2)
\nn \\
&-&3-\frac{\pi^2}{4}+\frac{1}{2}\Biggl[\Li{2}{\frac{1+c_1}{2}}+\Li{2}{\frac{1+c_2}{2}}+
\Li{2}{\frac{1+c}{2}}\Biggr]-\frac{1}{2}\ln^2(\frac{\epsilon''}{\epsilon}).
\ea

As the dominant (DL) contribution arises from the region of
intermediate state with a soft lepton in presence of photons, we
can generalize the above result including the emission of "hard"
internal and external photons. This can be realized by the choice
\ba \Delta_1\sim\Delta_2\sim \epsilon''. \ea In this case the
hadronic block remains the same as for the lowest order of PT.

In DL approximation we have \ba \delta^{DL}=\left
[L_t+\frac{1}{2}(L_1+L_2) \right ]\ln\frac{\epsilon''}{\epsilon}
-\frac{1}{2}\ln^2\frac{\epsilon''}{\epsilon}. \ea Up to now we
calculate the contribution of box diagram to the matrix element.
For the calculation of the asymmetry we must to take into account
imaginary pairs of the contribution. By general arguments of
analiticity  \cite{Eden} the matrix diagram with two-photon
exchange and insertion of vertex function with correction of
$\alpha^n$ order in DL approximation is proportional to $L_s
L_t^{2n-1}$.

It is naturally to expect (keeping in mind the arguments in favor
of exponentiation of soft photon emission contributions proved by
Yenne, Frautchi and Suura \cite{YFS}) that this result can be
generalized to all orders of PT: \ba
R=\frac{\textbf{Im}(M_{box}^*M_{born})^{(DL)+corr}}{\textbf{Im}(M_{box}^*M_{born})^{(DL)}}
&=&\biggl[\int\limits_0^{L_t} \frac{(1-c) \,\,\dd O''\,\,
\dd l}{2\pi(1-bc_1)(1-bc_2)}
\exp \left (\frac{\alpha}{\pi}\delta^{DL}\right )\biggr]
\nn \\
&\times&
\biggl[\int\limits_0^{L_t} \frac{(1-c) \,\,\dd O''\,\,
\dd l}{2\pi(1-bc_1)(1-bc_2)}\biggr]^{-1}.
\ea
We can use here
\begin{gather}
\dd O''=\frac{2\dd c_1 \dd c_2}{\sqrt{1-c_1^2-c_2^2-c^2+2cc_1c_2}}, \nn \\
\delta^{DL}=\frac{3}{4}L_tl+\frac{1}{8}l^2-\frac{7}{8}L_t^2+\frac{1}{4}(l-L_t)\left [\ln(1-bc_1)+\ln(1-bc_2)\right ],
\nn \\
l=\ln\frac{2(\epsilon'')^2(1-c)}{m^2}.
\end{gather}

The calculation in the lowest order of PT leads to \be R_{LO}
\approx 1-\frac{\alpha}{\pi}\frac{7 L_t^2}{24}. \label{eq:eqr} \ee
The numerical results for $R$ in the lowest order, $R_{LO}$, and
in higher orders, $R_\infty$, are presented in Fig. 5. One can see
they are sizable and should be taken into account.

\begin{figure}
\includegraphics[scale=1.]{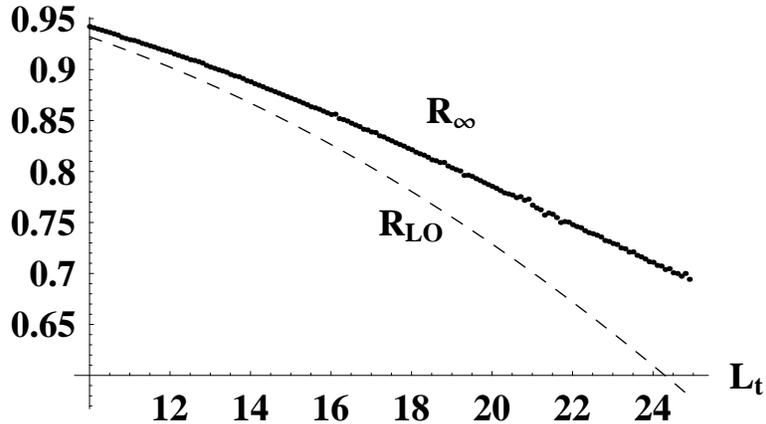}
\caption{ Numerical results for $R$, Eq. (\protect\ref{eq:eqr}).
$R_{LO}$ (dased line) is the result of the  calculation with RC in
lowest order of PT, $R_\infty$ (dotted line) when all orders of PT
are taken into account. We assume $c=\frac{1}{2}$.}
\end{figure}

The asymmetry at the lowest order of PT has been calculated in
previous papers, (see for instance \cite{AM04}).

Mass suppressed amplitudes connected with Higgs production and
decay in DL approximation were calculated in the paper
\cite{Fadin97}.

\section{Conclusion}

We have calculated higher order contributions to the asymmetry for elastic scattering of transversally polarized electrons on unpolarized protons. In particular we have shown that the double logarithmic corrections arising from the inelastic intermediate state in the two photon exchange amplitude can not be neglected.

\begin{figure}
\includegraphics[scale=1.]{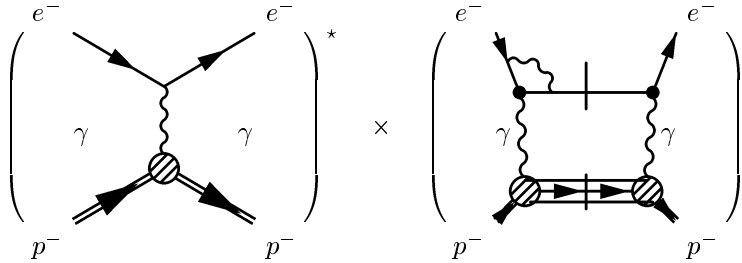}
\caption{Square box contribution to the asymmetry}
\label{fig_sq_box}
\end{figure}

\begin{figure}
\includegraphics[scale=1.]{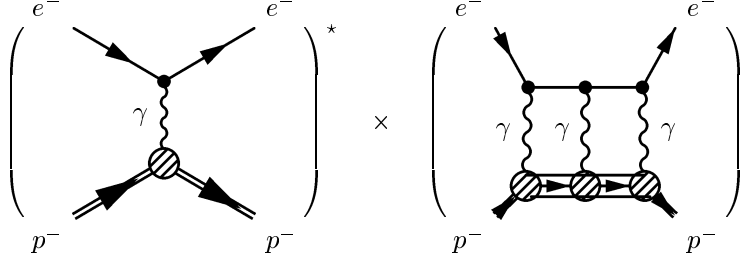}
\caption{Contribution to the asymmetry of the interference of the Born diagram with the two loop box diagram}
\label{tr_phot}
\end{figure}

The contribution to the imaginary part of the amplitude from the
square of the box diagram (Fig. \ref{fig_sq_box}) is of order
$(\alpha/\pi)^2L$, and can be omitted in DL approximation. This
holds also for the interference of the born diagram with the two
loop box diagram (Fig. \ref{tr_phot}) \cite{ax}.

We note that the limiting case $t\to 0$ can not be obtained using the
approach described above. Contrary to the hadronic block, where the $t\to 0$ limit
can be put smoothly, the leptonic block drastically depends on the parameter $(-t/m^2)$. For the $t=0$ case, the exact cancellation of RC related to the  internal emission of virtual and real photons takes place, as it was shown in Appendix D of \cite{BFKK}.

The presence of infrared singularities in RC to the impact factor of the electron was previously noted in the paper of one of us \cite{KLS}. In the present paper,  the way to eliminate such singularities is explicitly shown.

\section{Acknowledgments}

We are grateful for valuable remarks to the A. Kobushkin and D. Borisyuk.
Two of us (E.A.Kuraev, V.V. Bytev) are partially supported by the INTAS grant  05-1000008-8328
and MK-2952.2006.2.

\end{document}